# SNAP Telescope

M.Lampton[f], C.Akerlof[b], G.Aldering[a], R.Amanullah[c], P.Astier[d], E.Barrelet[d],
C.Bebek[a], L.Bergstrom[c], J.Bercovitz[a], G.Bernstein[e], M.Bester[f], A.Bonissent[g],
C.Bower[h], W.Carithers[a], E.Commins[f], C.Day[a], S.Deustua[i], R.DiGennaro[a], A.Ealet[g],
R.Ellis[j], M.Eriksson[c], A.Fruchter[k], J-F.Genat[d], G.Goldhaber[f], A.Goobar[c],
D.Groom[a], S.Harris[f], P.Harvey[f], H.Heetderks[f], S.Holland[a], D.Huterer[l], A.Karcher[a],
A.Kim[a], W.Kolbe[a], B.Krieger[a], R.Lafever[a], J.Lamoureux[a], M.Levi[a], D.Levin[b],
E.Linder[a], S.Loken[a], R.Malina[m], R.Massey[n], T.McKay[b], S.McKee[b], R.Miquel[a],
E.Mortsell[c], N.Mostek[h], S.Mufson[h], J.Musser[h], P.Nugent[a], H.Oluseyi[a], R.Pain[d],
N.Palaio[a,], D.Pankow[f], S.Perlmutter[a], R.Pratt[f], E.Prieto[m], A.Refregier[n], J.Rhodes[o],
K.Robinson[a], N.Roe[a], M.Sholl[f], M.Schubnell[b], G.Smadja[p], G.Smoot[f], A.Spadafora[a],
G.Tarle[b], A.Tomasch[b], H.von der Lippe[a], R.Vincent[d], J-P.Walder[a] and G.Wang[a]

[a] Lawrence Berkeley National Laboratory, Berkeley CA, USA
[b] University of Michigan, Ann Arbor MI, USA
[c] University of Stockholm, Stockholm, Sweden
[d] CNRS/IN2P3/LPNHE, Paris, France
[e] University of Pennsylvania, Philadelphia PA, USA
[f] University of California, Berkeley CA, USA
[g] CNRS/IN2P3/CPPM, Marseille, France
[h] Indiana University, Bloomington IN, USA
[i] American Astronomical Society, Washington DC, USA
[j] California Institute of Technology, Pasadena CA, USA
[k] Space Telescope Science Institute, Baltimore MD, USA
[l] Case Western Reserve University, Cleveland OH, USA
[m] CNRS/INSU/LAM, Marseille, France
[n] Cambridge University, Cambridge, UK
[o] NASA Goddard Space Flight Center, Greenbelt MD, USA
[p] CNRS/IN2P3/INPL, Lyon, France

## ABSTRACT

The SuperNova/Acceleration Probe (SNAP) mission will require a two-meter class telescope delivering diffraction limited images spanning a one degree field in the visible and near infrared wavelength regime. This requirement, equivalent to nearly one billion pixel resolution, places stringent demands on its optical system in terms of field flatness, image quality, and freedom from chromatic aberration. We discuss the advantages of annular-field three-mirror anastigmat (TMA) telescopes for applications such as SNAP, and describe the features of the specific optical configuration that we have baselined for the SNAP mission. We discuss the mechanical design and choice of materials for the telescope. Then we present detailed ray traces and diffraction calculations for our baseline optical design. We briefly discuss stray light and tolerance issues, and present a preliminary wavefront error budget for the SNAP Telescope. We conclude by describing some of tasks to be carried out during the upcoming SNAP research and development phase.

# 1. THE SNAP MISSION REQUIREMENTS

The SuperNova/Acceleration Probe (SNAP) is a planned satellite experiment designed to precisely measure the expansion history of the universe. The experiment is motivated by the remarkable discovery[1, 2] of an accelerating expansion rate. This acceleration suggests that the universe contains some form of cosmological constant, or dark energy. To effectively test models of the expansion, it is essential to compare accurate observational data against model predictions of the expansion rate as a function of lookback time, or equivalently as a function of redshift. Type Ia supernovae populate the observable universe and serve as accurately standardizable candles. Each measured supernova furnishes a redshift and a magnitude. The redshift is a measure of the expansion between its epoch and the present, while the magnitude is a measure of the elapsed time since the supernova exploded. Properly calibrated and sorted into systematic classes, a collection of a few thousand such supernovae spanning the redshift range $0<z<1.7$ will provide important new constraints on models of the universe and the dark energy that it contains. A description of the mission and its science is presented at the SNAP home page, http://snap.lbl.gov, and in companion papers by Aldering et al[3] and Kim et al[4].

The mission will be conducted by repeatedly scanning a 7.5 square degree zone near the north ecliptic pole for a 16 month period, and later conducting a similar study near the south ecliptic pole. During each study, scans will repeat with a four day cadence. Each scan will provide photometry in nine bands spanning the visible and near IR for all objects in the zone down to a faint magnitude limit of about AB=27.5. In this way, the light curve for each detected supernova will be determined in order to provide magnitude and classification data. During each four day period, time will also be taken to perform follow-up spectroscopy on each detected supernova near its maximum light to determine its redshift and obtain further classification data. To carry out these measurements, a large passively cooled multiband imager with approximately 600 million pixels will occupy much of the focal plane[5]. The focal plane is shared with a high-efficiency low-dispersion spectrometer[6] equipped with an integral field unit (IFU).

The requirements placed on the SNAP telescope derive directly from the science goals and the mission constraints. The wavelength coverage is determined by the need to measure a number of filter bands across the visible and near IR wavelength range, spanning roughly 0.35 microns to 1.7 microns, and to conduct low resolution spectroscopy of each supernova near maximum light to extract features allowing detailed classification. This requirement effectively rules out refracting optical trains, and drives the telescope towards all-reflective optics. The light gathering power is set by the need to discover distant supernovae early in their expansion phases and to permit accurate photometry and low resolution spectroscopy near maximum light. This requirement can be met with a minimum aperture of about two meters. Image quality is also a factor in determining signal to noise ratio (SNR) because of the effects of natural Zodiacal light and detector noise. For a two-meter aperture and one-micron wavelength, for example, the Airy disk size is 0.13 arcseconds FWHM and we intend to achieve angular resolution near the diffraction limit at wavelengths longward of one micron. To match this diffraction spot size to the size of typical silicon pixels (~10 microns) one must adopt an effective focal length of about 20 meters. This same focal length is also a good match in the near IR where wavelengths up to 1.7 microns are to be observed using typical HgCdTe detectors whose pixels are 18 microns in size[5]. Finally, a large field of view is needed for its multiplex advantage: a large number of sky pixels being observed in parallel contributes directly to the observing time per target for a given cadence and survey field size. Our mission constraints are met if this field of view is the order of one square degree, of which about 0.7 square degree will be instrumented by detector pixels. The ratio of working field area to diffraction patch area is about 800 million, comparable to the total number of detector pixels. By means of dithering we expect to recover photometric measurements good to a few percent accuracy. Undersampling, dithering, cosmic ray hits, and many other effects are included in the exposure time calculator developed by Bernstein[7].

The image quality of the telescope is driven in part by the SNR requirement, and also by the potential systematic supernova spectrum contamination by unwanted light from the supernova host galaxy. We have presently baselined a system Strehl ratio of 0.90 at one micron wavelength, corresponding to an RMS wavefront error (WFE) of 50 nm, or a Strehl ratio of 0.77 at the commonly used test wavelength of 0.633 microns.

Mission constraints and cost constraints are also factors that limit the size and configuration of the SNAP telescope. For example, a high Earth orbit is highly advantageous from the standpoint of achieving passive detector cooling. We plan to fly a three day orbit period with a perigee high enough to avoid the inner Van Allen belt whose energetic protons would otherwise seriously limit mission and detector lifetime. Launcher fairings for boosters having the requisite capability place significant limits on the combined length of the payload and spacecraft. We expect an overall payload length of about 6 meters and a payload diameter of about 2.5 meters will accommodate our two-meter aperture telescope.

Our primary science target fields are located near the north and south ecliptic poles where natural Zodiacal light is minimized for best near IR sensitivity. Observing these locations places the sun at nearly right angles to our view direction. In a high Earth orbit, a low orbital inclination serves to keep the Earth and moon also nearly at right angles to our view direction. We utilize this viewing geometry in several ways. First, the solar panels can be rigidly body-mounted on the sunward side of the spacecraft, which avoids the cost, failure modes, and structural flexibility of deployed panels. Second, the passive cooling radiator can be rigidly located on the antisunward side of the spacecraft, in permanent shadow. Third, the stray light baffling can be optimized for a limited range of solar roll and elevation angles, and for a limited range of Earth elevation angles. We plan to have the spacecraft perform 90 degree roll maneuvers every 3 months during the mission, to keep up with the mean ecliptic longitude of the sun. The detector array has a 90 degree roll symmetry that allows its photometric data acquisition to continue from season to season.

Prospective launch vehicles (Delta IV, ATLAS V, SeaLaunch) and payload fairing dimensions impose limits on the overall telescope size as well as its mass properties. Through a series of packaging exercises we have explored ways to fit the maximum length stray light baffle into available launch fairings, and find that a short optical package, ~3 m in length, can yield favorable stray light rejection when situated near the bottom of a tall outer baffle.

The launch environment imposes both stiffness and strength requirements on the payload. Vehicle aero-elastic stability concerns prescribe the needed payload stiffness in terms of minimum structural frequencies in the axial (~25 Hz) and lateral (~10 Hz) directions. The launch environment includes both quasi-steady and random acceleration events that are combined to establish the peak loads, or strength requirements for the payload. Preliminary estimates indicate the design loads will be the order of 12 Gs axial and 8 Gs in the lateral directions. The thermal environment will be 0°-40°C pre-launch, while the operational temperatures for the optical elements and structure will be actively controlled as dictated by the optics dimensional stability and instrument requirements.

## 2. DOWNSELECTION OF CONFIGURATION

A key parameter describing any telescope is its telephoto advantage, defined as the ratio of the system effective focal length (EFL) to the length of the optical package. If our EFL is about 20 meters and the package length is to be about 3 meters, we require a telephoto advantage of 6.7. Single mirror telescopes such as prime focus paraboloids or Schmidt cameras do not possess any useful telephoto advantage, and we do not consider them further. Two-mirror telescopes of the Gregorian, Cassegrain, or Ritchey-Chretien type can achieve any desired telephoto ratio, but have limited fields of view that fail to meet SNAP requirements, and in addition have seriously curved focal surfaces that complicate the use of large format detector systems. Field-widening lens groups and field-flattener lenses can be added, such as the Gascoigne corrector, but introduce serious chromatic aberrations that are unacceptable when spanning a wide wavelength range, and introduce potential radiation dose limitations which could be problematic in a long-life space mission.

Three-mirror telescopes free of refractive elements provide the solution SNAP needs. Early studies by Paul[8] and by Baker[9] sought field-widening achromatic correctors for large parabolic primary mirrors such as the Hale 5-meter telescope. Subsequent work by Angel et al[10] and McGraw et al[11] has demonstrated the feasibility of these configurations. An alternative approach, in which the primary mirror shape is regarded as a parameter freely adjustable along with the secondary and tertiary mirror curvatures and shapes, have been pursued by a number of authors. One such group of designs is the "three-mirror anastigmat" (TMA) family in which the astigmatism and field curvature are kept near zero during the optimization process.

Korsch[12] distinguished two types of TMA: those in his "Design I" group have a continually converging light beam which terminates on the focal surface with no intermediate focus and no defined exit pupil or beam waist. Those in his "Design II" group have an intermediate focus lying between the secondary and tertiary mirrors. This focus supplies a field stop location for stray light control. An image of the primary mirror is formed partway between the tertiary and the focal plane, offering a defined exit pupil which can serve as a Lyot stop or cold stop useful in managing infrared detector irradiance. Design I was explored further by Robb[13], Korsch[14], Epps & Takeda[15], Willstrop[16], Badiali and Amoretti[17], the LSST/DMT team[18], and others. Design II was explored further by Cook[19], Williams[20], and others, who developed off-axis eccentric field designs. The exceptionally good baffling offered by the integral field stop and the exit pupil Lyot stop has led to space flight TMAs for remote sensing that use this eccentric field layout. Two examples now in orbit are the Kodak IKONOS instrument[21] and the LANL/Sandia Multispectral Thermal Imager[22, 23, 24].

Korsch[25] meanwhile devised two more TMA implementations that possess the field stop and Lyot stop of his Design II yet offer an on-axis primary-secondary arrangement and a larger but annular field. These arrangements (see also Abel[26]) use a flat folding mirror that rotates the final focal surface away from the primary mirror axis, placing it to one side, thereby allowing free access to a large focal plane instrument. These two configurations are identical with regard to aberrations and speed, but differ in the location of the folding mirror. His "Configuration I" puts the tertiary mirror on the axis of the primary and secondary, and the folding mirror intercepts the light passing through the exit pupil en route to the focal surface. In this way the light from the secondary is not blocked by the detector instrumentation. In "Configuration II" the folding flat is an annular mirror located near the Cassegrain quasifocus, and redirects this intermediate image light towards the tertiary mirror relocated to one side of the primary axis. The central hole in this flat allows the tertiary light to reach the focal plane along a transverse axis just behind the primary. Both configurations are blind or heavily vignetted on axis -- Configuration I because the small folding flat blocks the center of the intermediate image, Configuration II because the hole in its annular flat cannot redirect that image center. But if the exit pupil is small and is located near this folding flat, the blind center of the telescope's field can block less than half of the total working field diameter.

With either configuration, the detector is moved away from the primary optical axis so that the detector no longer blocks any secondary light. For SNAP such an advantage is mandatory since our detector is larger than our secondary mirror. This configuration also provides a natural way to obtain passive detector cooling, since one side of the telescope will always face the antisunward direction. With Configuration II the total optical package length is reduced and the volume behind the primary mirror is more effectively utilized. We have adopted this general layout scheme for SNAP, and have fine-tuned it to meet our requirements for focal length, package length, vignetting, and field of view. A schematic view is shown in Figure 1.

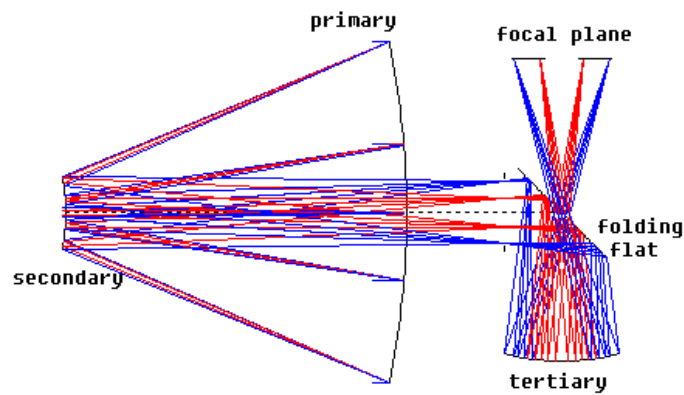

Fig. 1: SNAP optics layout. The entrance pupil is defined by the primary mirror. A field stop is located behind the primary mirror (vertical marks) for stray light control. The exit pupil is at the folding mirror.

# 3. OPTICAL CONFIGURATION

The SNAP detailed optical configuration has been selected by an iterative process involving exploring various alternative choices for focal length, working field coverage, and packaging constraints. The design parameters that control the aberrations are the three powered mirror curvatures, the three conic constants, and the three spacings (primary to secondary, secondary to tertiary, and tertiary to focal plane). These nine parameters are adjustable, subject to several constraints. The effective focal length (EFL), the primary aperture, and the focal plane diameter we regard as given. The element spacings and the consequent locations of the field stop and the exit pupil have to be carefully managed to achieve a satisfactory package and to place the exit pupil within the central hole of the folding mirror. The optimization process utilized a commercial ray tracer that incorporates a robust nonlinear least squares routine[27].

Because the TMA is blind on axis, it is of no particular value to base its optimization on an expansion of the paraxial or small-angle aberrations. Instead, a number of field points were chosen that populate the working field, ranging from 6 to 13 milliradians off axis angle, and the optimization proceeded using a total sum-of-squares demerit function. During each optimization, the element locations were held constant as a way of manually controlling the locations of the field stop and the cold stop. In this way, the Cassegrain quasifocus could be kept situated behind the primary mirror, very near the position of the folding flat mirror, so that the central blind spot (defined by cass focus light failing to be reflected from the flat) is well defined to lend a sharp transition between axial and lateral beams. The exit pupil is located in the center of the hole in the flat. The overall length of the optical train is 3.3 meters. Compared with the 21.66 meter effective focal length, this system has an effective telephoto advantage of about 6.5.

INDIVIDUAL COMPONENT DESCRIPTIONS

The optimized optical parameters are summarized in Table 1. The mirrors are pure conic sections of revolution having no polynomial terms. The use of higher polynomial terms has not yet been explored. The location of the vertex of each element is listed in a Cartesian (X,Z) coordinate system whose origin is the vertex of the primary mirror.

Table 1: Optical Surfaces and Locations

|  | Diameter, meters | Central hole, meters | Curvature, recip meters | Asphericity | Xlocation, meters | Zlocation, meters |
|---|---|---|---|---|---|---|
| Primary | 2.00 | 0.5 | -0.2037466 | -0.981128 | 0 | 0 |
| Secondary | 0.45 | none | -0.9099607 | -1.847493 | 0 | -2.00 |
| Folding flat | 0.66 x 0.45 | 0.19 x 0.12 | 0 | 0 | 0 | +0.91 |
| Tertiary | 0.68 | none | -0.7112388 | -0.599000 | -0.87 | +0.91 |
| Focal plane | 0.567 | 0.258 | 0 | 0 | +0.9 | +0.91 |

# 4. MECHANICAL CONFIGURATION

For a space mission it is vital to create a mechanical configuration that provides an extremely stable metering structure that maintains the optical element alignment during ground testing, launch, and orbit operations. The concept adopted for SNAP is to create three structural components that will be brought together during spacecraft/payload integration: a stiff low-precision outer baffle cylinder carrying the exterior solar panels and extensive thermal insulation; a stiff low-precision spacecraft bus structure that carries antennas, batteries, and other major spacecraft support components; and a stiff high-precision telescope structure comprising carbon-fiber metering elements, the kinematically-mounted mirrors, the instrumentation suite, and its own thermal control system. Fig. 2 (below) shows the overall payload and spacecraft layout, while Fig. 3 shows details of the secondary and tertiary metering structures.

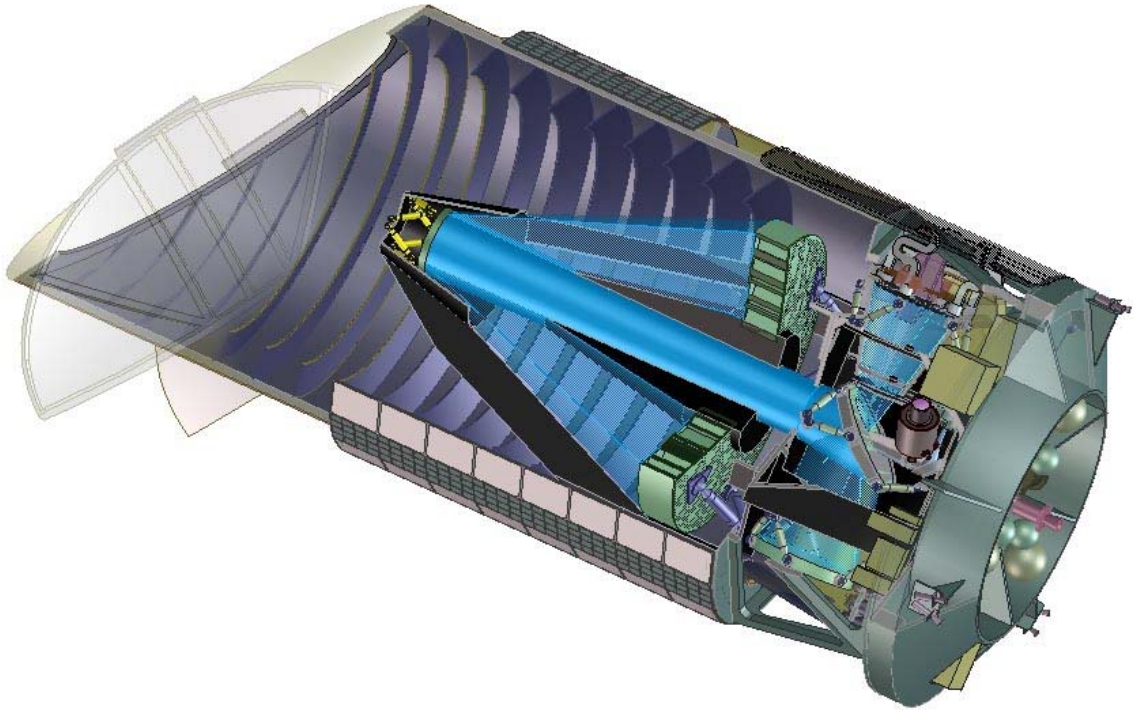

Fig. 2: Cutaway view of SNAP. The entire telescope telescope attaches to the spacecraft structure at right by means of bipods. The outer baffle, shown cut away, also attaches to the spacecraft structure by means of its separate supporting struts. A hinged split door, shown open in light gray, protects the cleanliness of the optics until on-orbit commissioning begins. Solar panels are fixed, not deployed.

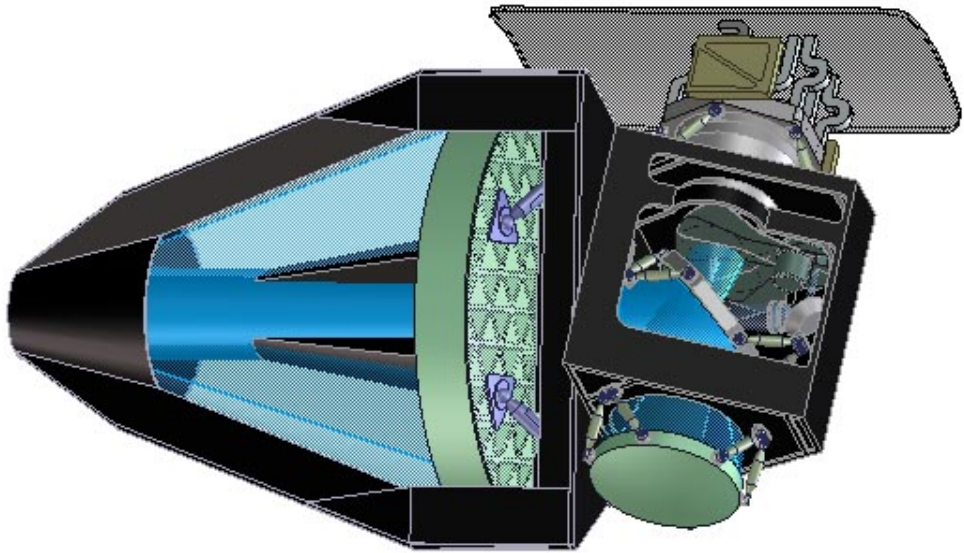

Fig. 3: Telescope metering structure (carbon fiber, shown in dark gray) provides precision control of optical element spacings and orientations. Forward of the primary mirror, the secondary is supported on adjusters within the secondary baffle. Aft of the primary mirror, the tertiary metering structure supports the folding flat, the tertiary, and the focal plane instrumentation. The passive radiator at top is thermally but not structurally linked to the focal plane instrumentation.

## 5. MATERIALS

Space-proven optical mirror technology is largely based on two approaches: open-back Schott Zerodur glass ceramic composite material and Corning ultra-low expansion ULE glass honeycomb structure. For SNAP either technology has sufficiently low coefficient of thermal expansion and sufficiently well proven manufacturing techniques. Studies are underway exploring the detailed fabrication and test flows using either process. Alternative materials, including various formulations of silicon carbide, are under study for other missions and may prove to be competitive for SNAP.

The metering structure will utilize a low-CTE carbon-fiber construction. In particular, the secondary support tripod will have to maintain the primary to secondary spacing accurate to a few microns. This tripod and the other major metering components will certainly require a dedicated active thermal control system. We anticipate the need for five-axis motorized adjustment for the secondary mirror during ground integration, on-orbit observatory commissioning, and occasionally during science operations. For this reason we plan to include a hexapod or other multi-axis positioner into the secondary support structure.

## 6. GEOMETRIC-OPTICS PERFORMANCE

The optical performance of our baseline optical telescope is fundamentally limited by aberrations and manufacturing errors at sufficiently short wavelengths, and by diffraction at long wavelengths. Accordingly, our expected performance figures divide into two areas: the geometrical ray traces that quantify the aberrations and the pupil diffraction studies. We summarize the key performance items in Table 2 and Fig. 4 below.

Table 2: Performance Summary

| | |
|---|---|
| Focal Length | 21.66 meters |
| Aperture | 2.0 meters |
| Final focal ratio | f/10.83 |
| Field | Annular, 6 to 13 mrad;  1.37 sq deg |
| RMS geometric blur | 2.8 microns, average 1 dimension |
| Central obstruction | 16% area when fully baffled |
| Vane obstruction | 8% area, tripod or quadrupod |

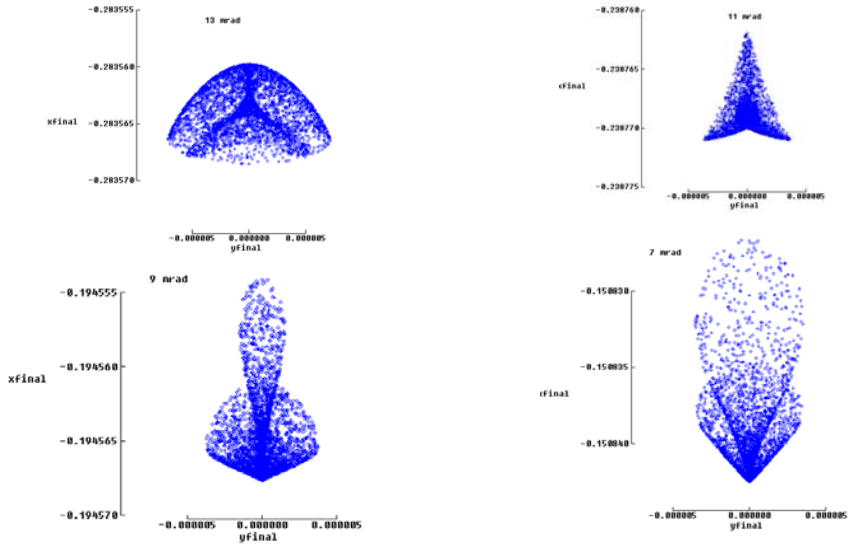

Fig. 4: Ray trace spot diagrams. Upper left: 13 mrad off axis; upper right 11 mrad; lower left 9 mrad; lower right 7 mrad. Tick marks are spaced 5 microns in the focal plane.

From spot diagrams at various off axis angles, we have compiled the statistics on the mean radial centroid of ray hits in the focal plane, and the second moments of the spot distributions. These are listed in the Table 3 in the form of the two orthogonal RMS breadths (radial RMS and tangential RMS) in columns 3 and 4. Combining these by their root-sum-square gives a two-dimensional measure of the spot size, listed in column 5 below as a linear focal plane dimension, and as an angular size on the sky in milliarcseconds in column 6. Finally, column 7 lists the effective one dimensional FWHM assuming a Gaussian conversion factor of 2.35 between 1-D RMS and 1-D FWHM. At the inner and outer radii of the image annulus, the FWHM becomes as large as 60 milli-arcsec, although in the midrange of the annulus it is smaller.

Table 3: Image Moments vs. Off Axis Angle

| OffAxis sin(theta) | Rfinal, microns | radial RMS, microns | tangentialRMS microns | TotalRSS, microns | TotalRSS, milliarcsec | FWHM, milliarcsec |
|---|---|---|---|---|---|---|
| 0.006 | 129122 | 3.32 | 1.60 | 3.69 | 34.88 | 57.97 |
| 0.007 | 150838 | 3.33 | 1.60 | 3.69 | 34.97 | 58.11 |
| 0.008 | 172649 | 3.18 | 1.59 | 3.56 | 33.65 | 55.92 |
| 0.009 | 194565 | 2.83 | 1.51 | 3.21 | 30.36 | 50.45 |
| 0.010 | 216600 | 2.28 | 1.37 | 2.66 | 25.17 | 41.84 |
| 0.011 | 238769 | 1.57 | 1.35 | 2.07 | 19.60 | 32.57 |
| 0.012 | 261086 | 1.18 | 1.89 | 2.23 | 21.09 | 35.05 |
| 0.013 | 283565 | 2.09 | 3.23 | 3.85 | 36.41 | 60.51 |
|  |  |  | AVERAGE= | 3.12 | 29.52 | 49.05 |

Distortion is another fundamental optical aberration, but unlike the other Seidel aberrations distortion does not impact the SNR nor does it directly impact the detection of supernovae. It does however cause the loci of scanned field objects to depart from parallel tracks in the focal plane, and does complicate the weak lensing science. Any mapping of the celestial sphere onto a plane surface causes some distortion owing to the differing metrics of curved and flat spaces. In our baseline design, we have disregarded distortion as a driver, in order to use all available design variables to maximize the working field of view and minimize the net geometrical blur. It is nonetheless important to explore the resulting distortion quantitatively. The TMA distortion is axisymmetric owing to the symmetry of the unfolded (powered) optical train, and in polar coordinates any off-axis angle maps onto a single focal plane radius independent of azimuth angle. The distortion is therefore purely radial. Table 4 lists the radial distance of an off axis field point as a function of the sine of the off axis angle, and the departure from proportionality to the sine of that angle.

Table 4: Radial Distortion

| sin(theta) | R,microns | LinModel | Diff,microns |
|---|---|---|---|
| 0.006 | 129122 | 129960 | -838 |
| 0.007 | 150838 | 151620 | -782 |
| 0.008 | 172650 | 173280 | -630 |
| 0.009 | 194565 | 194940 | -373 |
| 0.010 | 216600 | 216600 | 0 |
| 0.011 | 238769 | 238260 | 509 |
| 0.012 | 261085 | 259920 | 1165 |
| 0.013 | 283565 | 259920 | 1983 |

From Table 4 it is seen that the TMA distortion is of the pincushion type, having increased magnification towards the extremity of the field. Compared to a linear mapping of sin(theta) onto focal plane radius, the distortion amounts to about two percent.

## 7. PUPIL DIFFRACTION

For a star at infinity and a telescope focussed at infinity, the pupil diffraction pattern is computed using the Fraunhofer formalism, and the focal plane irradiance is simply the square of the modulus of the two dimensional Fourier transform of the pupil. For quantitative studies of our expected point spread function and our diffracted light background, we have computed this irradiance function for a variety of prospective pupils. Figure 5 below shows this irradiance in a two-dimensional logarithmic format. The vertical scale shows the extent of five orders of magnitude of irradiance. The pupil, shown at the right, has a two meter aperture, three tripod legs of 50mm width, and a central obstruction 0.7m in diameter. The assumed wavelength is 1.0 microns. The six spikes and the central Airy disk patterns are evident.

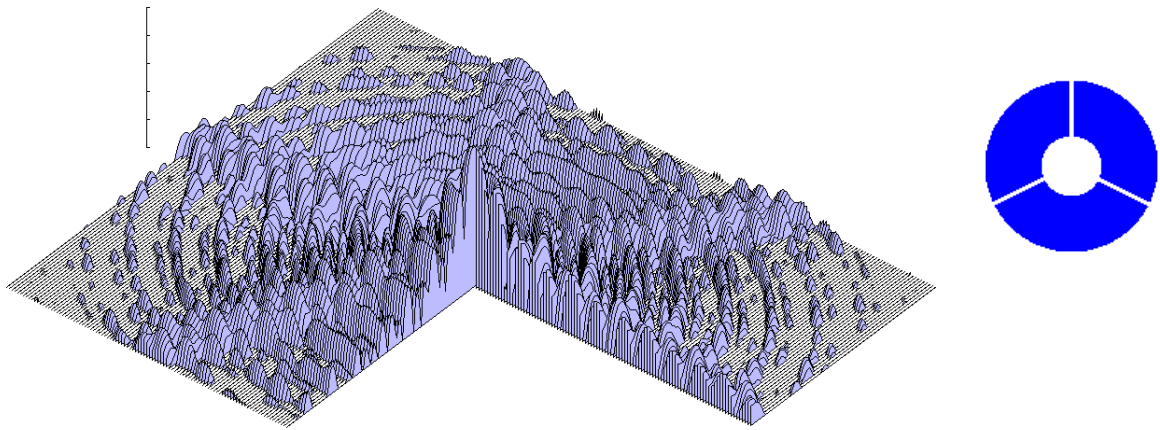

Figure 5: focal plane irradiance defined by diffraction of a monochromatic incoming plane wave, 1.0 microns wavelength, through the pupil shown at right. Vertical scale (upper left) shows logarithmic five-decade range of irradiance. Horizontal span is 5 x 5 arcseconds with steps of 0.023 arcsecond per image slice. The threefold symmetry of the pupil causes the six diffraction spikes evident in the figure.

## 8. STRAY LIGHT

A comprehensive stray light control plan has been developed for SNAP. Our goal is to keep all stray light sources far below the natural Zodiacal irradiance level as seen at the focal plane. The primary concern is of course sunlight scattered past the forward edge of the outer light baffle. This will require a minimum of two successive forward edges, since the light diffracted past a single edge would exceed the allowable irradiance at the primary mirror, assuming typical mirror scattering values. Another concern, during portions of the orbit where the fully illuminated Earth is seen, is scattered Earth light. When fully illuminated, the Earth stands opposite to the sun and the tall interior side of the outer baffle tube receives Earthshine. We have devised a baffle angle strategy that will help minimize this radiation seen at the primary mirror (see Figure 6 below). The blades are angled downward, so that even at the lowest Earth elevation, Earthshine reaches only their upper surfaces, while the primary mirror can see only their dark lower surfaces. In this way, a minimum of two scatters is needed for Earth light to reach the primary. Additional stray light occurs from the moon, stars, etc, and is being quantitatively tracked as our design process continues.

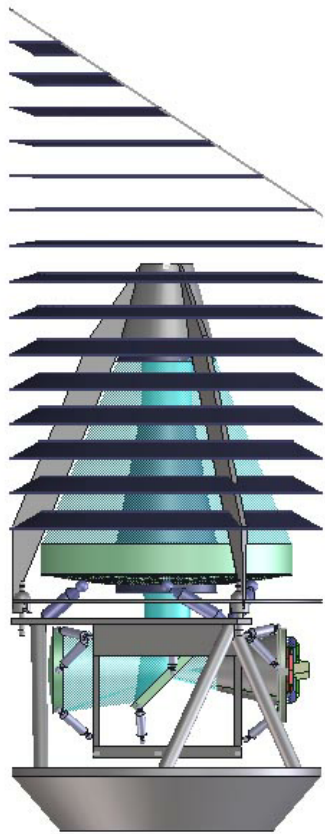

Fig. 6: Schematic treatment of the outer baffle interior vane arrangement. Sunlight is incident from the left, where the height of the baffle and its angled forward edge maintains the baffle interior in darkness. Earthshine is at times incident from the right, however, and therefore the vane angles require particular attention so that the lower vane surfaces are not illuminated by the Earth.

## 9. TOLERANCES

A tolerance budget has begun with a group of exploratory studies of the sensitivity of the geometrical spot size to variations in element curvatures, shapes, locations, and orientations. Initial assessment of these calculations shows that by far the single most critical parameters are the primary mirror curvature and the spacing between primary and secondary mirrors. This result is expected owing to the fast (f/1.2) primary mirror. A two-micron displacement of the secondary piston, or a two-micron displacement in the virtual image created by the primary mirror, increases the RMS geometrical blur by about 3 microns.

Similarly, a 15-micron lateral displacement or a 15-microradian tilt of the secondary mirror causes a corresponding 3-micron growth in the RMS geometrical blur. The other optical elements are far less critical because the magnifications from those surfaces is smaller by an order of magnitude.

The baseline SNAP telescope includes on-orbit mechanical adjustments that permit the relocation and reorientation of the secondary mirror, and possibly the tertiary mirror as well, to optimize image quality. By means of these adjustments we anticipate accommodating small shifts in any of the optical elements locations and orientations.

## 10. WAVEFRONT ERROR BUDGET

The departure of any surface from its nominal mathematical conic section, or the misplacement or misorientation of any of the surfaces, causes a wavefront error and a degraded image quality. One measure of this degradation is the telescope's Strehl ratio, which is the peak monochromatic image irradiance divided by the theoretical peak irradiance for the ideal diffraction limited image. Strehl ratio can be converted into RMS wavefront error (RMS WFE) through Marechal's relation. To achieve a system Strehl ratio of 0.77 at 0.633 microns wavelength, the total WFE must not exceed 50 nm rms. This allowed WFE can be apportioned into individual contributions for planning purposes. Such an apportionment is listed in Table 5.

Table 5:  Wavefront Error Budget, RMS nanometers, fully adjusted & collimated

| | |
|---|---|
| Primary figure | 33 nm |
| Secondary position | 5 nm zeroable by telecommand |
| Secondary orientation | 5 nm zeroable by telecommand |
| Secondary figure | 5 nm |
| Folding flat position | 5 nm |
| Folding flat orientation | 5 nm |
| Folding flat figure | 5 nm |
| Tertiary position | 5 nm zeroable by telecommand |
| Tertiary orientation | 5 nm zeroable by telecommand |
| Tertiary figure | 5 nm |
| Detector position | 10 nm |
| Detector orientation | 10 nm |
| Detector flatness | 10 nm |
| Manager's reserve | 18 nm |
| TOTAL root sum square | 43 nm |

SNAP is presently embarked on a two year research and development program during which a number of trade studies will be conducted. Many of these trades involve the telescope: its exact dimensions, its selection of materials, its manufacturing and testing, and the overall SNAP integration plan. Among the telescope tasks is a study to define the most appropriate budget for WFE terms. Since these include a variety of manufacturing and testing error contributions, such a budget will have to be worked with a full understanding of the manufacturing and test procedures, including in particular the means of evaluating data taken in a one-gravity environment with respect to the performance expected in a zero-gravity environment.

## 11. CONCLUSIONS

We have presented an overview of the requirements and status of a telescope design for the planned SNAP mission. The optical, mechanical, and thermal studies and analyses conducted to date indicate that this telescope is manufacturable and testable using proven techniques. Upcoming work during the SNAP research and development phase will further refine these concepts.

## ACKNOWLEDGMENTS

This work was supported by the Director, Office of Science, of the U.S. Department of Energy under Contract No. DE-AC03-76SF00098.

**Contact:** mlampton@SSL.berkeley.edu; tel (USA) 510-642-3576; Space Sciences Lab, University of California, Berkeley CA 94720 USA.